\documentclass[]{eptcs}
\providecommand{\event}{ICLP 2019 DC} 
\usepackage{breakurl}             
\usepackage{underscore}           
\usepackage{listings}
\title{Imperative Program Synthesis from Answer Set Programs}
\author{Sarat Chandra Varanasi
\institute{University of Texas at Dallas, Richardson, TX, United States}
\email{sxv153030@utdallas.edu}}

\usepackage{mathrsfs}
\def\titlerunning{Imperative Program Synthesis from Answer Set Programs}
\def\authorrunning{Sarat Chandra Varanasi}
\begin{document}
\maketitle

\begin{abstract}
Our research concerns generating imperative programs from Answer Set Programming Specifications. ASP is highly declarative and is ideal for writing specifications. Further with \textit{negation-as-failure} it is easy to succinctly represent combinatorial search problems. We are currently working on synthesizing imperative programs from ASP programs by turning the negation into useful computations. This opens up a novel way to synthesize programs from executable specifications. 
\end{abstract}
\section{Introduction and Problem Description}
      Our research area is broadly in Program Synthesis from Logical Specifications. Currently we are working on synthesizing imperative code from Answer Set Programming specifications. ASP enables one to succinctly model various planning and combinatorial search problems \cite{KahlGelfond}. These specifications heavily use negation-as-failure which is the cornerstone feature of ASP. What ASP represent are executable specifications. We want to transform them into meaningful and efficient imperative programs. We target imperative programs because our overarching goal is to be able to synthesize concurrent data structures from concurrency specifications written in ASP. 
      \par Concurrent Programs are very difficult to design and debug. Our idea is to partially evaluate the ASP concurrency specification with respect to the operational semantics provided by s(ASP) \cite{arias2018constraint, marple2017computing} and obtain concurrent imperative code adhering to the specification.
      \par Concurrent In ASP, specification and computation are synonymous which is highly desirable to rapidly prototype systems \cite{eiter2005data}. However the implementations of ASP are quite complex  \cite{eiter1997disjunctive, GebserLMPRS18}. On the other hand, imperative programs reflect a simpler operational model of computation making them easier to trace and in general enjoy faster run-times than their declarative program counterparts. This is because imperative programs often represent computationally faster implementations of a specification without directly being concerned about the specification. This makes deriving imperative programs from ASP specifications highly desirable. Our main insight is to employ the recently-developed operational semantics for ASP realized in the s(ASP) system \cite{marple2017computing, arias2018constraint}. s(ASP) is a query-driven answer set programming system which computes the (partial) stable models of a query against an answer set program without grounding the program first. Traditional implementations of ASP are based on grounding the program to its propositional equivalent and then use a procedure similar to SAT solvers to compute the \textit{answer sets} \cite{GebserLMPRS18}. Using the operational semantics of s(ASP), one can follow how the stable models are computed for a query in a step-by-step manner. This provides us the right playground, to extract an imperative program while simplifying all the machinery that makes s(ASP) work.

\section{Background}
    Program synthesis has been tackled in various ways. One of the earliest works are related to theorem-proving \cite{manna1971towards}. More recently synthesis has been reduced to verification \cite{srivastava2010program, solar2006combinatorial}. The work of Srivastava et al. \cite{srivastava2010program} assumes a logical input-output specification and generates finite syntactic program structures (candidate programs) and then use Hoare-logic \cite{hoare1969axiomatic} to verify that the syntactic structures indeed satisfy the input-output specifications. To that end, verification is reduced to constraint-solving. We take a different direction in our research. The correct computational specifications are already given. All we need to do is to sieve out an imperative program. To our knowledge, no one else has attempted transforming ASP specifications to imperative programs till date.
    
    \medskip\noindent{\textbf{Prominent Features of s(ASP):}} We next summarize the salient features of the s(ASP) system that are relevant to our work here. 
         The operational semantics of s(ASP) is a Prolog style execution of an issued query adhering to the stable model semantics. The query triggers a search  where all clauses 
         that are consistent are resolved and backtracking happens when  inconsistent goals are encountered. 
         All the terms encountered in the consistent goals constitute the ``partial'' stable model associated with the query. No grounding of clauses is needed. The procedure to decide which clauses are consistent or inconsistent is quite sophisticated. At a simple level, the current resolvent is invalid when during expansion it leads to a stable model that has both a goal and its negation. This is in fact the way how stable models are constructed in the propositional variant of s(ASP), namely Galliwasp \cite{marple2012goal}. 
         Note that the execution algorithm of s(ASP) relies on \textit{corecursion} \cite{simon2006coinductive} to handle even loops \cite{marple2017computing}, but this feature is not as important for the work reported here. 

         \medskip\noindent{\textbf{Handling negation with Dual rules:}} In s(ASP) negated goals are executed through dual rules. The dual rule of a predicate $p$ systematically negates the literals in a rule body defining $p$. Dual rules of all predicates together with the rules in the original program represent its \textit{completion}\cite{lloyd2012foundations}.
         For instance, the dual rule of the predicate $p$ with the following definition is shown below:
         \begin{lstlisting}[basicstyle=\selectfont\ttfamily, mathescape]
            %definition of p    %dual of p/0
            p :- not q.         not p :- np 1, np2
            p :- r, not s.      
                                %negating of not q yields q
                                np1 :- q.
                                
                                %negating r, not s yields 
                                %disjunction $(not \ r) \lor s$
                                
                                np2 :- not r.
                                np2 :- s.
          
         \end{lstlisting}
         
         \medskip\noindent{\textbf{Forall mechanism:}} We have just shown dual rules for propositions which are simple enough. However, writing dual rules for predicates is more involved. One complication is due to implicit quantifiers in predicate rules. For a rule such as:
         \begin{lstlisting}[basicstyle=\selectfont\ttfamily] 
                  p(X) :- q(Y), r(X).
          \end{lstlisting}
          $X$ is universally quantified over the whole formula, whereas $Y$ is existentially quantified in the body only. Therefore, negating $p(X)$
          results in $Y$ being universally quantified in the body as follows:
          \begin{lstlisting}[basicstyle=\selectfont\ttfamily]
                not p(X) :- np1(X), np2(X).
                np1(X) :- forall(Y, not q(Y)).
                np2(X) :- not r(X).
          \end{lstlisting}
         Notice that the universal quantifier for $Y$ in $not \ q(Y)$ is enclosed within a $forall(...)$.
         The $forall$ represents a proof procedure which runs through all values in the domain of $Y$ and verifies if the goal enclosed within it is satisfied. In our example, the $forall$ checks that $not \ q(Y)$ holds for all values of in the domain of $Y$. No such mechanism existed in prior Prolog based systems. It is this $forall$ mechanism coupled with dual rules that we make extensive use of in our synthesis procedure. Because they treat negated predicates $constructively$ \cite{marple2017computing}, we turn them into computations in our synthesis procedure.
         
\section{Research goal}
   As mentioned before, our main goal is to synthesize efficient concurrent programs. Writing programs that are \textit{correct-by-construction} is highly desirable because it automatically alleviates the burden of testing, debugging and formal verification. Our research is centered around the s(ASP) system which is still in experimental stage. The s(ASP) system provides a valid operational semantics. This enables a user to issue a query and get answer sets relevant to his query. This is to be contrasted with traditional ASP solvers which ground the entire program and solve answer sets using some form of SAT-solving \cite{GebserLMPRS18}. Another goal is to speedup the s(ASP) system by fine-tuning the goal-directed execution. Currently the top-down query evaluation in s(ASP) needs further improvement\footnote{As stated earlier, s(ASP) is experimental yet} in performance. This is due to the way constraints are handled by the s(ASP) engine. This is addressed in Section 6.  
   
\section{Current status of the research}
     We are able to synthesize sequential combinatorial algorithms such as graph-coloring, n-queens etc. We are progressing towards dynamic domains such as planning problems. For instance, modelling a concurrent linked list is no different than blocks-world planning. The blocks-world problem can be described in the $Action$ language  $\mathscr{A}$ \cite{KahlGelfond}. A concurrent list also represents a dynamic domain. For example, updating the nodes in a list are no different than placing one block over another. The updates are subject to having ``locks''. These restrictions represent executability conditions in Action languages. We have written the entire specification of a concurrent list execution. What remains is to come up with an appropriate procedure to extract the corresponding imperative algorithm.
     
\section{Preliminary results}
     Right now we can synthesize imperative programs for a class of datalog answer set programs. Further all the variables in the ASP program must be safe and come from finite domains. To properly motivate our work we give the following example.
     
      Consider the ASP program that finds the maximum of $n$ numbers, that is naturally specified in ASP as shown below: 
       \lstset{
       language=Prolog
        }
    \begin{lstlisting}[basicstyle=\selectfont\ttfamily]
           max(X) :-  num(X), not smaller(X). 
           smaller(X) :- num(X), num(Y), X < Y.
    \end{lstlisting}
    
    \par $num(X)$ provides the domain of numbers over which the input values range. $smaller(X)$ defines when a number $X$ is dominated by another number $Y$. $max(X)$ gives the definition for $X$ to be the maximum. The main predicate from where the computation to decide whether a given number $X$ is the maximum begins with $max(X)$. The negation of $smaller(X)$ can be translated to a $forall$ as shown below:
    \begin{lstlisting}[basicstyle=\selectfont\ttfamily]
    not smaller(X) :- forall(Y, not (num(X), num(Y), X < Y)).
    \end{lstlisting} 
    
    Assuming that $num(X)$ and $num(Y)$ are true, the negation only applies to $X < Y$. Therefore, 
    replacing the definition of $not$ $smaller(X)$ in $max(X)$ gives us the following: 
    \begin{lstlisting}[basicstyle=\selectfont\ttfamily]
         max(X) :-  forall(Y, num(X), num(Y), not (X < Y)). 
    \end{lstlisting} 
    
    The $forall$ definition of $max(X)$ makes apparent the operational flavor involved in finding the maximum of $n$ numbers: enumerate all numbers $Y$ and compare them with $X$. If $X$ is not smaller than any $Y$ ie. $X < Y$ is false for all $Y$, then $X$ is maximum. The $forall$ can be translated to a \textit{for-loop} in imperative languages if the domains of the variables involved in the scope of $forall$ are finite. From an Answer Set Programming point of view, $max(X)$ is present in the answer set if the $forall$ succeeds (if $X$ is the maximum), otherwise it is not present. The abstract code synthesized through program tranformation, thus looks like: 
      \begin{lstlisting}[basicstyle=\selectfont\ttfamily]
                def max(x):
                    for y in num:
                        if x < y:
                            return False
                    return True
       \end{lstlisting}
   We have submitted an entire paper explaining this idea in detail to the LOPSTR 2019 conference. The main insight is that the $forall$ mechanism from s(ASP) can be directly translated into \textit{for-loops} (atleast for finite domains).

\section{Open Challenges and Expected Achievements}
The work reported here is the first step towards obtaining efficient implementation from  high level specifications. Our eventual goal is to specify algorithms (e.g., inserting, deleting, and checking for membership of an element) for concurrent data structures (lists, queues, trees, heaps, etc.) as answer set programs and automatically be able to obtain the efficient, imperative versions of those algorithms. Achieving such a goal requires making  extensive use of partial evaluation and semantics-preserving re-arrangement of goals in the current resolvent. It is part of our future work.
Currently our synthesis procedure does not handle global constraints (head-less rules in ASP).
For example our work supports synthesis of the graph coloring problem written in the form:
\begin{lstlisting}[basicstyle=\selectfont\ttfamily]
color(X, C) :- node(X), color(C), not another_color(X,C), 
                  not conflict(X,C).
another_color(X, C) :- node(X), color(C), color(C1), C != C1,
                  color(X, C1).
conflict(X, C) :- node(X), color(C), node(Y), X != Y, edge(X, Y), 
                  color(X, C), color(Y, C).
\end{lstlisting}
But not the form written with global constraints as follows:

\begin{lstlisting}[basicstyle=\selectfont\ttfamily]
color(X, C) :- node(X), color(C), not another_color(X, C).
 
another_color(X, C) :- node(X), color(C), node(Y), 
                       X != Y, color(Y, C).
 
:-  color(X, C), color(Y, C), edge(X, Y).
      \end{lstlisting}

Essentially, we should be able to take constraints and move them up in the resolvent to the earliest point where this constraint can be executed.
    
      \lstset{language=Prolog}
      Handling global constraints is closely related to dynamic consistency checking (DCC) in goal-directed answer set programs \cite{marple2014dynamic, marple2012goal}. DCC ensures only the necessary constraints, that are relevant to a rule and the possible bindings variables could take,  are checked along with the rule-body. All other constraints which are not relevant to the rule can be ignored. This has been addressed for ground answer set programs. Our future work focuses on providing DCC for predicates. This would also enable synthesis of programs specifying dynamic domains such as data structures, and planning problems. 
\bibliographystyle{eptcs}
\bibliography{generic}
\end{document}